\documentclass[aps,apl,twocolumn,reprint,superscriptaddress,amsmath,amssymb]{revtex4-1}

\usepackage{xcolor}
\usepackage{graphicx}
\usepackage{multirow}
\usepackage{booktabs}
\usepackage{colortbl}
\usepackage{tabularx}
\usepackage{hhline}
\usepackage{upgreek}
\usepackage[
			colorlinks=true,
			linkcolor=blue,
			citecolor=blue,
			filecolor=green
			pagebackref=true,
			]{hyperref} 

\newcommand\etal{\textit{et al}}

{\setlength{\tabcolsep}{0.15cm}

\begin{document}

\title{Exciton-exciton annihilation in hBN}

\author{A. Plaud}
\affiliation{Laboratoire d'Etude des Microstructures, ONERA-CNRS, Universit\'e Paris-Saclay, 29 avenue de la division Leclerc, BP 72, 92322 Ch\^atillon Cedex, France}
\affiliation{Groupe d'Etude de la Mati\`ere Condens\'ee, UVSQ-CNRS, Universit\'e Paris-Saclay, 45 avenue des Etats-Unis, 78035 Versailles Cedex, France}
\author{L. Schu\'e}
\affiliation{Laboratoire d'Etude des Microstructures, ONERA-CNRS, Universit\'e Paris-Saclay, 29 avenue de la division Leclerc, BP 72, 92322 Ch\^atillon Cedex, France}
\affiliation{Groupe d'Etude de la Mati\`ere Condens\'ee, UVSQ-CNRS, Universit\'e Paris-Saclay, 45 avenue des Etats-Unis, 78035 Versailles Cedex, France}
\author{K. Watanabe}
\affiliation{National Institute for Materials Science, 1-1 Namiki, Tsukuba 305-0044, Japan}
\author{T. Taniguchi}
\affiliation{National Institute for Materials Science, 1-1 Namiki, Tsukuba 305-0044, Japan}
\author{A. Loiseau}
\email{annick.loiseau@onera.fr}
\affiliation{Laboratoire d'Etude des Microstructures, ONERA-CNRS, Universit\'e Paris-Saclay, 29 avenue de la division Leclerc, BP 72, 92322 Ch\^atillon Cedex, France}
\author{J. Barjon}
\email{julien.barjon@uvsq.fr}
\affiliation{Groupe d'Etude de la Mati\`ere Condens\'ee, UVSQ-CNRS,  Universit\'e Paris-Saclay, 45 avenue des Etats-Unis, 78035 Versailles Cedex, France}


\begin{abstract}
Known as a prominent recombination path at high excitation densities, exciton-exciton annihilation (EEA) is evidenced in bulk hexagonal boron nitride (hBN) by cathodoluminescence at low temperature. Thanks to a careful tune of the the exciton density by varying either the current or the focus of the incident electron beam, we could estimate an EEA rate of 2$\times$10$^{-6}$ cm$^{3}$.s$^{-1}$ at $T=10$ K, the highest reported so far for a bulk semiconductor. Expected to be even stronger in nanotubes or atomic layers, EEA probablly contributes to the luminescence quenching observed in low-dimensionality BN materials.

\end{abstract}


\maketitle

2D materials are of great interest as novel physical properties arise from their highly anisotropic structure. They can be thinned down to the monolayer and various layered material can be stacked to create an artificial Van der Waals heterostructure with controlled and multifunctional properties. Hexagonal boron nitride, an indirect wide bandgap semiconductor (\textgreater 6 eV), is one of the key materials in this topic. It has been demonstrated to be the best insulating material for improving electron mobility in graphene \cite{Dean2010}, and revealing close-to-homogeneous excitonic linewidths in transition metal dichalcogenides \cite{Cadiz2017} as a well as a suitable dielectric spacer in devices \cite{Geim2013}. Nevertheless, some properties of hBN remains to be  fully understood, starting with the most prominent of them which is its luminescence efficiency in the deep UV range (215 nm). Recently we measured a $\sim$50\% internal quantum yield for the radiative recombinations of indirect excitons in hBN and evidenved its $\sim$300 meV binding energy \cite{Schue2018}. Such tightly-bound excitons brings the radiative efficiency of an indirect bandgap semiconductor to the level of a direct one. Such unique and outstanding luminescence features clearly deserve further investigations.

Up to now, the luminescence properties have been poorly investigated, due to instrumental difficulties in conducting suitable photoluminescence experiments in the UV-C range (4.43 - 12.4 eV). An alternative approach is to perform cathodoluminescence (CL).   In such a way, Watanabe \etal. could already point out the strong luminescence of hBN as well as a lasing effect in their pioneering work on single crystals\cite{Watanabe2004}. In the Fig.5 of ref \cite{Watanabe2004}, below the stimulation threshold, one could notice a sublinear dependence of the luminescence intensity with increasing electron beam current. This is unusual since a linear dependence is generally expected for free exciton recombinations in semiconductors \cite{Schmidt1992}. A sublinearity might be caused by exciton-exciton annihilation (EEA), a type of Auger recombinations for excitons, which is evidenced in this work.

For affecting the performances of transistor devices operating at high power, lasers and solar cells, Auger recombinations of free charge carriers have been studied extensively since the early times of semiconductor physics (see Ref. \cite{Landsberg1992} for a review). The efficiency drop of GaN light-emitting diode (LED) at high current is a current example where Auger recombinations are still under discussion \cite{Iveland2013}. Auger effects also occur with free excitons and are known to prevent reaching the high exciton densities required for the investigation of many-body effects. For instance, EEA is detrimental for Bose Einstein condensation (BEC) of excitons in copper oxyde crystals \cite{Schwartz2012}. Enhanced at low dimension, EEA limits the luminescence efficiency in a large panel of nanomaterials such as 2D layers \cite{Kumar2014,Yuan2015}, carbon nanotubes \cite{Wang2004} and quantum dots \cite{Klimov2000}. 

In this letter, taking advantage of our recent effort for achieving measurements of the luminescence efficiency using CL \cite{Schue2018}, we present a quantitative study of the exciton-exciton annihilation (EEA) in high quality hBN crystals. Two kinds of experiments are performed. In the first one, hBN intrinsic luminescence intensity is classically studied as a function of the electron beam current to evidence the EEA effect in hBN. In the second experiment, the intensity is analyzed as a function of the excitation surface, providing a first estimate of the EEA rate in hBN.

Experiments are performed on a hBN single crystal grown with the high-pressure and high-temperature (HPHT) method \cite{Taniguchi2007a}. Such crystals are recognized as of the highest cristallinity and purity available today \cite{Schue2016}. The cathodoluminescence spectra were acquired on the best regions of the sample, selected by CL imaging, with a JEOL7001F field-emission-gun scanning electron microscope (SEM) coupled to a Horiba Jobin-Yvon detection system as described in Schu\'e \etal.  \cite{Schue2016}. Thanks to the intensity calibration of the setup with a reference deuterium lamp, the absolute CL intensity is expressed in photons/s and opens the possibility to evaluate the luminescence efficiency \cite{Schue2018}. The sample is mounted on a Gatan cooling stage for a CL analysis close to liquid helium temperature, the sample holder being at 10 K. In order to avoid charging effects during CL analysis, a 5 nm semi-transparent gold film was evaporated on the crystal. All results were obtained using a 5 keV  incident electron beam.

\begin{figure}[h!]
 \begin{center}
 \includegraphics[scale=1]{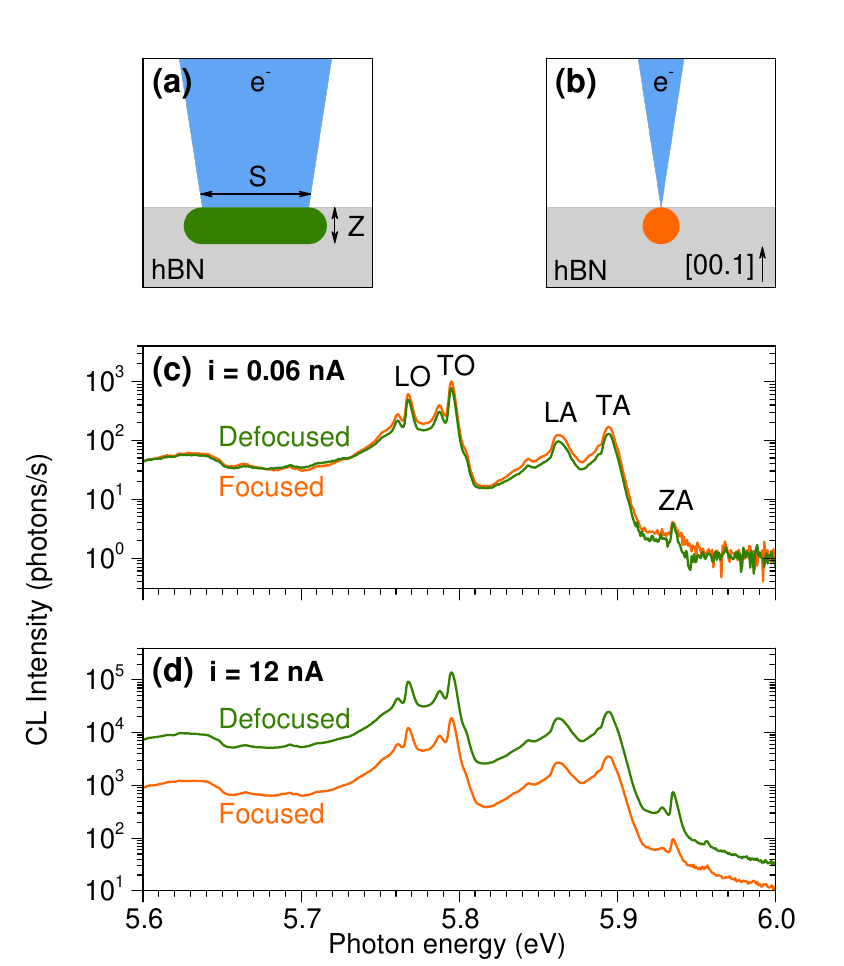}
 \caption{Effect of the excitation density on the radiative recombination intensity of free excitons in hBN at T = 10 K. Schematic diagram of defocused (a) and focused (b) electron beams exciting the sample. S denotes the irradiated surface area and z the electron penetration depth. CL spectra are compared when using defocused (S = 49 $\mu$m$^2$) or focused beams at a current (c) 0.06 nA and (d) 12 nA.}
 \label{F1}
 \end{center}
\end{figure}

Exciton-exciton annihilation (EEA) being dependent on the exciton density, one has to control the excited volume in order to investigate this phenomenon. At a given incident electron energy in a SEM, the exciton density can be tuned using either the beam current or the beam focusing as shown in the early works of Casey \etal. \cite{Casey1971} and sketched in Figure \ref{F1}(a) and (b). For current-dependent experiments, the beam current was varied from $i=$ 0.06 nA up to 12 nA, as measured with a Faraday cup. The other option is to defocus the electron beam in order to increase its impact surface on the sample. This is done here by overfocusing the electron beam \textit{i.e.} by decreasing the objective-lens focal length. The irradiated surface area S is measured thanks to the carbon contamination marks visible in secondary electron images. The focused beam diameter is typically 1.8 - 5 nm within the used current range. In defocused conditions the beam diameter was increased up to 14.5 $\mu$m. The beam defocusing increases the excited surface by almost 8 orders of magnitudes, which provides an efficient way to dilute the exciton density.

hBN CL spectra are displayed in Figure \ref{F1}. The near band edge luminescence spectrum exhibits five peaks at energies 5.935, 5.894, 5.864, 5.795 and 5.768 attributed to the radiative recombinations of the indirect free exciton, assisted by ZA, TA, LA, TO, LO phonons respectively \cite{Cassabois2016}. No peak broadening could be observed, even at highest excitation densities, which indicates the absence of significant heating of the exciton gas. The sample luminescence exhibits low defect-related luminescence: no D lines around 5.5 eV related to extended structural defects \cite{Pierret2014} could be detected, and deeper defects only give rise to a weak luminescence at 4.1 eV as seen in the 200-400 nm spectrum (see Supplementary Material A).

At low beam current (Fig.\ref{F1}(c)), the beam focusing is of no influence on the luminescence intensity. CL spectra are almost superimposable, meaning that the luminescence is not affected by the exciton density. At high current (Fig.\ref{F1}(d)), focusing the electron beam leads to a decrease of the CL intensity by an order of magnitude, while the other spectral features remain unchanged. This evidences a nonradiative recombination process which occurs at high exciton densities. Exciton-exciton annihilation is proposed here to account for the observed luminescence quenching and briefly described in what follows.

The exciton-exciton annihilation arises from an inelastic collision between two excitons, writing $X+X \rightarrow X$. It appears as a type of Auger process, where one of the excitons transfers its energy ($\approx$ 6  eV) to the other one, the excess of energy being further dissipated non-radiatively through multi-phonon emissions. The EEA can be described by the following rate equation:

\begin {equation}
\frac{dn}{dt}=g-\frac{n}{\tau}-\frac{\gamma n^2}{2} 
\label {eq1}
\end {equation}

\noindent where $n$ is the density of excitons (cm$^{-3}$), $g$ is the generation rate per volume unit (cm$^{-3}$.s$^{-1}$), $\tau$ is the exciton lifetime without EEA effect (s) achieved at low excitation and $\gamma$ is the EEA rate (cm$^{3}$.s$^{-1}$). The $1/2$ factor on  $\gamma$ accounts for the fact that the EEA leaves one remaining exciton over the two initial ones.

EEA is usually studied thanks to time resolved photoluminescence (TRPL) for various laser fluences. This has not been done for hBN yet but TRPL profiles can be found in the literature. They show a strong discrepancy in the temporal decay of the exciton luminescence: single exponential \cite {Watanabe2011} and multi-exponential decays \cite {Cao2013,Cassabois2016} were both reported. This is understood in this work as due to different excitation densities, a single exponential decay being the signature of low excitation conditions. The solution of equation (\ref{eq1}) in pulsed regime indeed unifies these apparently contradictory results, as shown in the Supplementary Material B. A free exciton lifetime of 600 ps at 8 K is extracted from the single exponential decay reported in Ref. \cite {Watanabe2011}. The result was obtained from a hBN single crystal also grown with the HPHT method under the same growth condition to our samples. We will then assume that $\tau$ = 600 ps under low excitation in the hBN crystal investigated here.

In the present CL experiment, EEA arises under a continuous electron-beam excitation provided in a SEM. In such a steady-state regime, the solution of the rate equation (\ref{eq1}) 
is expressed as

\begin {equation}
n=\frac{1}{\gamma \tau}(\sqrt{1+2\gamma \tau^{2}g}-1)
\label {eq2}
\end {equation}

The CL intensity is directly related to the amount of excitons radiatively recombinating: $I_{CL} = nV/\tau_{rad}$, where V (cm$^3$) is the volume occupied by the exciton gas, and $\tau_{rad}$ (s) the exciton radiative lifetime. By keeping all others experimental parameters fixed, the luminescence intensity as a function of the beam current, $i$, writes

\begin {equation}
I_{CL}(i)=A(\sqrt{1+Bi}-1)
\label {eq3}
\end {equation}

\noindent where the $A$ and $B$ factors are constants (see Supplementary Material C for their complete expressions). The classical behaviour is obtained at weak excitations $I_{CL}(i)\propto i$, while at strong excitations, the luminescence intensity undergoes a sublinear dependence $I_{CL}(i)\propto \sqrt{i}$ due to EEA.

\begin{figure}[h!]
 \begin{center}
 \includegraphics[scale=1]{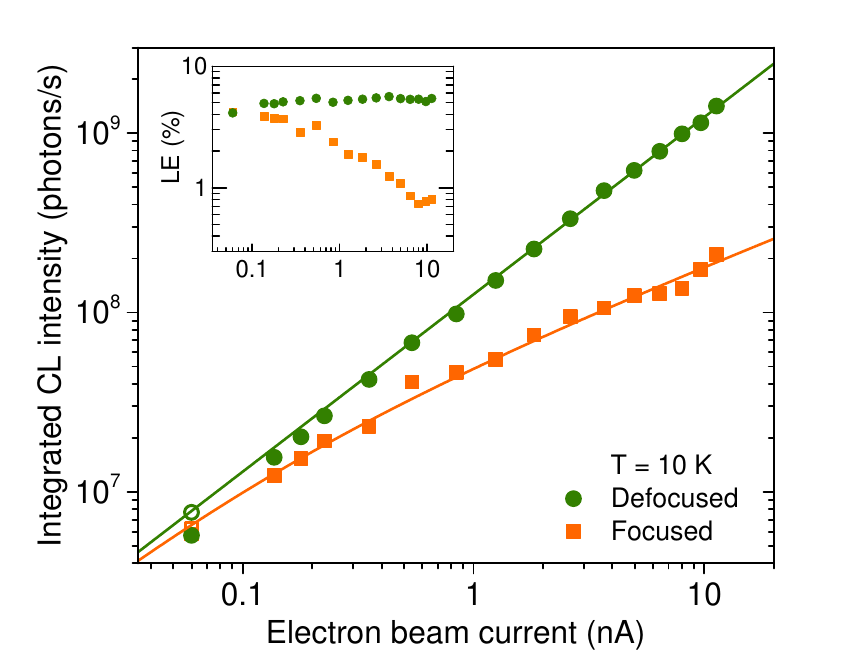}
 \caption{Integrated intensity of the free exciton CL (209-217 nm) as a function of the electron beam current. Circles are measured for a defocused beam ($S =$ 49 $\mu m^2$) and squares for a focused one as depicted in Fig.\ref{F1}(a) and (b). Data were collected for increasing current steps and ended with a repeated point at low current (open symbols) to check the absence of degradation. Solid lines correspond to fits described in the text. Inset: luminescence efficiency (LE).}
 \label{F2}
 \end{center}
\end{figure}

The results are plotted in Fig. (\ref{F2}) showing the CL intensity as a function of the electron beam current. A diluted excitation is first obtained using a defocused beam, such as the impact surface is kept at 49 $\mu$m$^2$ during the experiment. In these conditions of a diluted exciton gas, we observe the CL intensity is proportional to the excitation current, as shown by the unity-slope of the linear fit. This result indicates that many-body interactions such as EEA do not occur, excitons being too far apart from each other to interact. The luminescence efficiency appears almost constant at about 5\% over the studied beam current range, consistently with the value found at 5 kV in our previous work, performed at low current, \textit{i.e.} in the absence of EEA \cite{Schue2018}.

Much higher exciton densities are obtained when the electron beam is focused on the sample surface. In this case, Fig.\ref{F2} shows that the CL intensity exhibits a square root dependence at high currents, providing a strong evidence that the recombination process is based on two-particle collisions. This bimolecular process is non-radiative, the LE drops by an order of magnitude at 12 nA, where EEA appears as the main recombination channel for excitons. 

In the case of a focused beam excitation, the free exciton luminescence intensity is well described by Eq. (\ref{eq3}) as shown by the solid lines in Fig. 2 obtained with $A=1.4\times10^{7}$ s$^{-1}$ and $B=20$ nA$^{-1}$. However, the exciton diffusion length in hBN remains unknown (undoubtedly larger that the focused beam diameter), the effective volume $V$ occupied by excitons is therefore difficult to estimate as well as the EEA rate $\gamma$.

On the contrary, with a sufficiently defocused electron beam, \textit{i.e.}\ when the beam impact surface area is much larger than the exciton diffusion length, the excitation volume $V$ can be considered to be proportional to the impact surface area $S$. Here the volume was calculated assuming a cylindrical geometry $V= Sz$, where the depth of excitation $z = 308$ nm was obtained from Monte Carlo simulations at 5 keV (Sup. Mat. D) and $S$ was measured thanks to the contamination marks on SEM images.

\begin{figure}[h!]
 \begin{center}
 \includegraphics[scale=1]{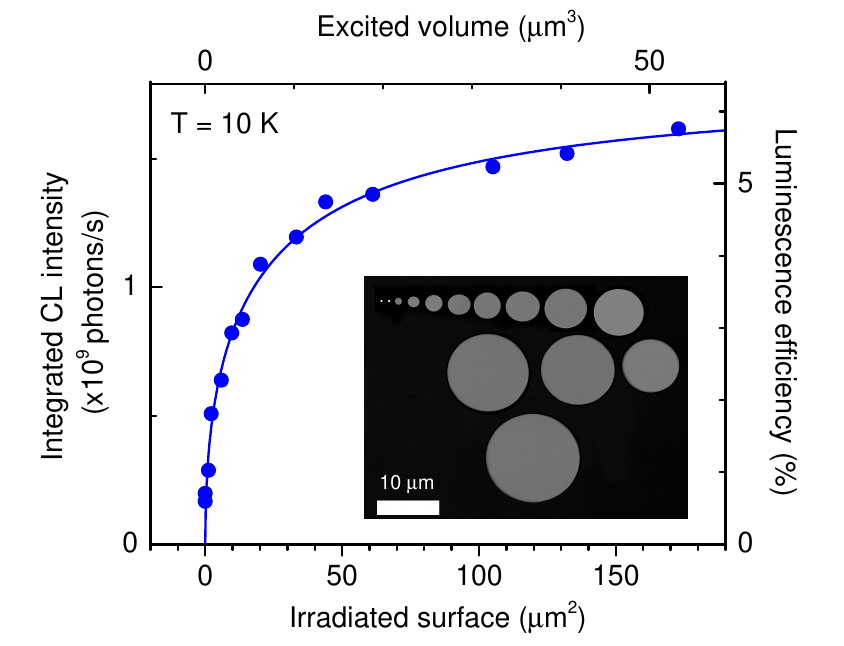}
 \caption{CL intensity of free exciton recombinations in hBN (integrated from 209 to 217 nm) as a function of the excited surface or volume with $i$=12 nA. The solid line is the best fit of $I_{CL}(V)$ according to Eq.\ref{eq4} giving $I_{max}=1.8 \times 10^9$ photons.s$^{-1}$ and $V_c=15.9$ $\mu$m$^3$. An SEM image of the contamination marks used to measure the irradiated surface area $S$ is shown in inset.}
 \label{F3}
 \end{center}
\end{figure}

Fig.\ref{F3} shows the results obtained in volume-dependent experiments at fixed current. As the electron beam is progressively defocused, we observe that the luminescence efficiency increases up to an order of magnitude. We find again a maximum LE of 5 \% as in the previous current-dependent experiment. Increasing the impact surface dilutes the exciton gas, the EEA becomes weak and the LE reaches its maximum. 

Eq. \ref{F2} also provides a description of the CL experiment as a function of the excitation volume at a fixed current, which expresses as
\begin {equation}
I_{CL}(V)= I_{max} \left (\sqrt{1+\frac{2V_c}{V}}-1\right )\frac{V}{V_c}
\label {eq4}
\end {equation}

\noindent where $I_{max}$ is the maximum CL intensity recorded in the limit of low exciton densities. $V_c=\gamma G\tau^2$ is a characteristic volume at which EEA becomes significant. Compared to $I_{max}$ obtained for a diluted exciton gas, the CL intensity drops by 27\% at $V_c$ and by 50\% at $V_c/4$ . Interestingly, $I_{max}$ and $V_c$ can be determined independently from the experimental data. The best fit of Fig.\ref{F3} according to Eq.\ref{eq4} gives $I_{max}=1.8 \times 10^9$ photons.s$^{-1}$ and $V_c=15.9$ $\mu$m$^3$. In this experiment, the generation rate $G$ is $1.9 \times 10^{13}$ s$^{-1}$ as deduced from the current and acceleration voltage used for the electron beam (see Supplementary Material C). With a lifetime of $\tau=600$ ps \cite{Watanabe2011}, the order of magnitude found for the EEA rate is $2 \times 10^{-6}$ cm$^3$.s$^{-1}$.

A characteristic exciton density for the occurence of the EEA can also be inferred from Fig.\ref{F3}. At $V_c$, the exciton density is $n_c=(\sqrt{3}-1)G\tau/V_c$. Then the EEA starts to occur (27\% of luminescence losses) at a typical density of excitons equal to $5.3 \times 10^{14}$ cm$^{-3}$ , which corresponds to an average exciton-exciton distance of 120 nm. It is important to remind that excitons are extremely compact in hBN. The in-plane extent of the exciton Bohr radius is only a few lattice constants \cite{Paleari2018}. The EEA process in hBN probably involves an exciton diffusion step as discussed in WSe$_2$ \cite{Mouri2014a}, black phosphorus \cite{Surrente2016} and organic semiconductors \cite{Engel2006, Ma2012, Shaw2008}. TRPL experiments would be needed to obtain a more detailed understanding of the EEA mechanism in hBN.

To our knowledge, the highest EEA rate in bulk semiconductors was obtained in Cu$_2$O crystals, in the 0.5-4$\times$10$^{-7}$ cm$^3$.s$^{-1}$ range at low temperature \cite{Yoshioka2010,Snoke2014}. The 2.2$\times$10$^{-6}$ cm$^{3}$.s$^{-1}$ EEA rate obtained for hBN clearly exceeds the Cu$_2$O case, so that hBN appears as the bulk semiconductor crystal where EEA is the strongest. The fundamental optical constants of hBN should then be measured at low exciton density. For applications to UV light sources\cite{Watanabe2009a}, EEA has also to be avoided to preserve the very bright hBN luminescence and specific designs will probably have to be developped.

At low dimensionality, EEA is generally enhanced \cite{Kumar2014, Wang2004, Klimov2000}. This is well illustrated in the experimental works of Yuan \etal.\cite{Yuan2015}, where the EEA rate recorded for a single layer (1L) WS$_2$ is two orders of magnitude larger than for 3L WS$_2$.  A few theoretical works have pointed out the relationship between EEA and exciton properties. In bulk crystals, it has been proposed that the EEA rate scales with the exciton Bohr radius $a$, $\gamma \propto 1/a^2$ \cite{Snoke2014}. In 1D systems such as carbon nanotubes, the EEA rate is expected to depend  on the bandgap, $E_g$, and exciton binding energy, $E_b$, giving $\gamma \propto (E_b/E_g)^3$ \cite{Wang2006}. Though theoretical works would clearly deserve to be completed, it seems clear that EEA is generally enhanced for excitons of high binding energy.

The 0.3 eV binding energy of the excitons responsible for the bulk hBN luminescence \cite{Schue2018} is predicted to increase to 2.1 eV for BN monolayers and more than 3 eV in BN nanotubes \cite{Wirtz2006}. As a consequence, excitons in 2D and 1D BN probably face giant EEA effects. This might be a reason explaining why no free exciton luminescence could be detected so far, neither for 2D hBN flakes thinner than 6L \cite{Schue2016}, nor for single-wall boron nitride nanotubes \cite{Pierret2014}.

As a summary, we have evidenced for the first time exciton-exciton annihilation in bulk hBN single crystals with CL experiments. The quantitative evaluation of the hBN EEA rate has been achieved by using a controlled defocusing of the electron beam. The order of magnitude found for EEA in hBN appears as the strongest reported so far in bulk semiconductors. The brightness promises of hBN for deep UV light-sources then faces an intrinsic limitation with a non radiative EEA process.  Expected to be even stronger in BN nanotubes or atomic layers, EEA should be also considered to explain the origin of their luminescence quenching.\\

\begin{acknowledgments}

The authors would like to thank professor Xavier Marie for fruitful discussions, C. Vilar for technical support and the French National Agency for Research (ANR) for funding this work under the project GoBN (Graphene on Boron Nitride Technology), Grant No. ANR-14-CE08-0018. The research leading to these results has also received funding from the European Commission H2020 Future and Emerging Technologies under grant agreements no. 696656 GrapheneCore1 and grant no. 785219 GrapheneCore2. Growth of hexagonal boron nitride crystals was supported by the Elemental Strategy Initiative conducted by the MEXT, Japan and the CREST (JPMJCR15F3), JST.
\end{acknowledgments}
\section*{Additional information}
The following article has been submitted to Applied Physics Letters. After it is published, it will be found at https://aip.scitation.org/journal/apl.

\newpage

\bibliography{hBN}

\begin{thebibliography}{31}%
\makeatletter
\providecommand \@ifxundefined [1]{%
 \@ifx{#1\undefined}
}%
\providecommand \@ifnum [1]{%
 \ifnum #1\expandafter \@firstoftwo
 \else \expandafter \@secondoftwo
 \fi
}%
\providecommand \@ifx [1]{%
 \ifx #1\expandafter \@firstoftwo
 \else \expandafter \@secondoftwo
 \fi
}%
\providecommand \natexlab [1]{#1}%
\providecommand \enquote  [1]{``#1''}%
\providecommand \bibnamefont  [1]{#1}%
\providecommand \bibfnamefont [1]{#1}%
\providecommand \citenamefont [1]{#1}%
\providecommand \href@noop [0]{\@secondoftwo}%
\providecommand \href [0]{\begingroup \@sanitize@url \@href}%
\providecommand \@href[1]{\@@startlink{#1}\@@href}%
\providecommand \@@href[1]{\endgroup#1\@@endlink}%
\providecommand \@sanitize@url [0]{\catcode `\\12\catcode `\$12\catcode
  `\&12\catcode `\#12\catcode `\^12\catcode `\_12\catcode `\%12\relax}%
\providecommand \@@startlink[1]{}%
\providecommand \@@endlink[0]{}%
\providecommand \url  [0]{\begingroup\@sanitize@url \@url }%
\providecommand \@url [1]{\endgroup\@href {#1}{\urlprefix }}%
\providecommand \urlprefix  [0]{URL }%
\providecommand \Eprint [0]{\href }%
\providecommand \doibase [0]{http://dx.doi.org/}%
\providecommand \selectlanguage [0]{\@gobble}%
\providecommand \bibinfo  [0]{\@secondoftwo}%
\providecommand \bibfield  [0]{\@secondoftwo}%
\providecommand \translation [1]{[#1]}%
\providecommand \BibitemOpen [0]{}%
\providecommand \bibitemStop [0]{}%
\providecommand \bibitemNoStop [0]{.\EOS\space}%
\providecommand \EOS [0]{\spacefactor3000\relax}%
\providecommand \BibitemShut  [1]{\csname bibitem#1\endcsname}%
\let\auto@bib@innerbib\@empty
\bibitem [{\citenamefont {Dean}\ \emph {et~al.}(2010)\citenamefont {Dean},
  \citenamefont {Young}, \citenamefont {Meric}, \citenamefont {Lee},
  \citenamefont {Wang}, \citenamefont {Sorgenfrei}, \citenamefont {Watanabe},
  \citenamefont {Taniguchi}, \citenamefont {Kim}, \citenamefont {Shepard},\
  and\ \citenamefont {Hone}}]{Dean2010}%
  \BibitemOpen
  \bibfield  {author} {\bibinfo {author} {\bibfnamefont {C.~R.}\ \bibnamefont
  {Dean}}, \bibinfo {author} {\bibfnamefont {A.~F.}\ \bibnamefont {Young}},
  \bibinfo {author} {\bibfnamefont {I.}~\bibnamefont {Meric}}, \bibinfo
  {author} {\bibfnamefont {C.}~\bibnamefont {Lee}}, \bibinfo {author}
  {\bibfnamefont {L.}~\bibnamefont {Wang}}, \bibinfo {author} {\bibfnamefont
  {S.}~\bibnamefont {Sorgenfrei}}, \bibinfo {author} {\bibfnamefont
  {K.}~\bibnamefont {Watanabe}}, \bibinfo {author} {\bibfnamefont
  {T.}~\bibnamefont {Taniguchi}}, \bibinfo {author} {\bibfnamefont
  {P.}~\bibnamefont {Kim}}, \bibinfo {author} {\bibfnamefont {K.~L.}\
  \bibnamefont {Shepard}}, \ and\ \bibinfo {author} {\bibfnamefont
  {J.}~\bibnamefont {Hone}},\ }\href {\doibase 10.1038/nnano.2010.172}
  {\bibfield  {journal} {\bibinfo  {journal} {Nat. Nanotechnology}\ }\textbf
  {\bibinfo {volume} {5}},\ \bibinfo {pages} {722} (\bibinfo {year}
  {2010})}\BibitemShut {NoStop}%
\bibitem [{\citenamefont {Cadiz}\ \emph {et~al.}(2017)\citenamefont {Cadiz},
  \citenamefont {Courtade}, \citenamefont {Robert}, \citenamefont {Wang},
  \citenamefont {Shen}, \citenamefont {Cai}, \citenamefont {Taniguchi},
  \citenamefont {Watanabe}, \citenamefont {Carrere}, \citenamefont {Lagarde},
  \citenamefont {Manca}, \citenamefont {Amand}, \citenamefont {Renucci},
  \citenamefont {Tongay}, \citenamefont {Marie},\ and\ \citenamefont
  {Urbaszek}}]{Cadiz2017}%
  \BibitemOpen
  \bibfield  {author} {\bibinfo {author} {\bibfnamefont {F.}~\bibnamefont
  {Cadiz}}, \bibinfo {author} {\bibfnamefont {E.}~\bibnamefont {Courtade}},
  \bibinfo {author} {\bibfnamefont {C.}~\bibnamefont {Robert}}, \bibinfo
  {author} {\bibfnamefont {G.}~\bibnamefont {Wang}}, \bibinfo {author}
  {\bibfnamefont {Y.}~\bibnamefont {Shen}}, \bibinfo {author} {\bibfnamefont
  {H.}~\bibnamefont {Cai}}, \bibinfo {author} {\bibfnamefont {T.}~\bibnamefont
  {Taniguchi}}, \bibinfo {author} {\bibfnamefont {K.}~\bibnamefont {Watanabe}},
  \bibinfo {author} {\bibfnamefont {H.}~\bibnamefont {Carrere}}, \bibinfo
  {author} {\bibfnamefont {D.}~\bibnamefont {Lagarde}}, \bibinfo {author}
  {\bibfnamefont {M.}~\bibnamefont {Manca}}, \bibinfo {author} {\bibfnamefont
  {T.}~\bibnamefont {Amand}}, \bibinfo {author} {\bibfnamefont
  {P.}~\bibnamefont {Renucci}}, \bibinfo {author} {\bibfnamefont
  {S.}~\bibnamefont {Tongay}}, \bibinfo {author} {\bibfnamefont
  {X.}~\bibnamefont {Marie}}, \ and\ \bibinfo {author} {\bibfnamefont
  {B.}~\bibnamefont {Urbaszek}},\ }\href {\doibase 10.1103/PhysRevX.7.021026}
  {\bibfield  {journal} {\bibinfo  {journal} {Phys. Rev. X}\ }\textbf {\bibinfo
  {volume} {7}},\ \bibinfo {pages} {021026} (\bibinfo {year}
  {2017})}\BibitemShut {NoStop}%
\bibitem [{\citenamefont {Geim}\ and\ \citenamefont
  {Grigorieva}(2013)}]{Geim2013}%
  \BibitemOpen
  \bibfield  {author} {\bibinfo {author} {\bibfnamefont {A.~K.}\ \bibnamefont
  {Geim}}\ and\ \bibinfo {author} {\bibfnamefont {I.~V.}\ \bibnamefont
  {Grigorieva}},\ }\href {\doibase 10.1038/nature12385} {\bibfield  {journal}
  {\bibinfo  {journal} {Nature}\ }\textbf {\bibinfo {volume} {499}},\ \bibinfo
  {pages} {419} (\bibinfo {year} {2013})}\BibitemShut {NoStop}%
\bibitem [{\citenamefont {{Schue}}\ \emph {et~al.}(2018)\citenamefont
  {{Schue}}, \citenamefont {{Sponza}}, \citenamefont {{Plaud}}, \citenamefont
  {{Bensalah}}, \citenamefont {{Watanabe}}, \citenamefont {{Taniguchi}},
  \citenamefont {{Ducastelle}}, \citenamefont {{Loiseau}},\ and\ \citenamefont
  {{Barjon}}}]{Schue2018}%
  \BibitemOpen
  \bibfield  {author} {\bibinfo {author} {\bibfnamefont {L.}~\bibnamefont
  {{Schue}}}, \bibinfo {author} {\bibfnamefont {L.}~\bibnamefont {{Sponza}}},
  \bibinfo {author} {\bibfnamefont {A.}~\bibnamefont {{Plaud}}}, \bibinfo
  {author} {\bibfnamefont {H.}~\bibnamefont {{Bensalah}}}, \bibinfo {author}
  {\bibfnamefont {K.}~\bibnamefont {{Watanabe}}}, \bibinfo {author}
  {\bibfnamefont {T.}~\bibnamefont {{Taniguchi}}}, \bibinfo {author}
  {\bibfnamefont {F.}~\bibnamefont {{Ducastelle}}}, \bibinfo {author}
  {\bibfnamefont {A.}~\bibnamefont {{Loiseau}}}, \ and\ \bibinfo {author}
  {\bibfnamefont {J.}~\bibnamefont {{Barjon}}},\ }\href@noop {} {\bibfield
  {journal} {\bibinfo  {journal} {ArXiv e-prints}\ } (\bibinfo {year}
  {2018})},\ \Eprint {http://arxiv.org/abs/1803.03766} {arXiv:1803.03766
  [cond-mat.mtrl-sci]} \BibitemShut {NoStop}%
\bibitem [{\citenamefont {Watanabe}\ \emph {et~al.}(2004)\citenamefont
  {Watanabe}, \citenamefont {Taniguchi},\ and\ \citenamefont
  {Kanda}}]{Watanabe2004}%
  \BibitemOpen
  \bibfield  {author} {\bibinfo {author} {\bibfnamefont {K.}~\bibnamefont
  {Watanabe}}, \bibinfo {author} {\bibfnamefont {T.}~\bibnamefont {Taniguchi}},
  \ and\ \bibinfo {author} {\bibfnamefont {H.}~\bibnamefont {Kanda}},\ }\href
  {\doibase 10.1038/nmat1134} {\bibfield  {journal} {\bibinfo  {journal} {Nat.
  Materials}\ }\textbf {\bibinfo {volume} {3}},\ \bibinfo {pages} {404}
  (\bibinfo {year} {2004})}\BibitemShut {NoStop}%
\bibitem [{\citenamefont {Schmidt}\ \emph {et~al.}(1992)\citenamefont
  {Schmidt}, \citenamefont {Lischka},\ and\ \citenamefont
  {Zulehner}}]{Schmidt1992}%
  \BibitemOpen
  \bibfield  {author} {\bibinfo {author} {\bibfnamefont {T.}~\bibnamefont
  {Schmidt}}, \bibinfo {author} {\bibfnamefont {K.}~\bibnamefont {Lischka}}, \
  and\ \bibinfo {author} {\bibfnamefont {W.}~\bibnamefont {Zulehner}},\ }\href
  {\doibase 10.1103/PhysRevB.45.8989} {\bibfield  {journal} {\bibinfo
  {journal} {Phys. Rev. B}\ }\textbf {\bibinfo {volume} {45}},\ \bibinfo
  {pages} {8989} (\bibinfo {year} {1992})}\BibitemShut {NoStop}%
\bibitem [{\citenamefont {Landsberg}(1992)}]{Landsberg1992}%
  \BibitemOpen
  \bibfield  {author} {\bibinfo {author} {\bibfnamefont {P.~T.}\ \bibnamefont
  {Landsberg}},\ }\href {\doibase 10.1017/CBO9780511470769} {\emph {\bibinfo
  {title} {Recombination in Semiconductors}}}\ (\bibinfo  {publisher}
  {Cambridge University Press},\ \bibinfo {year} {1992})\BibitemShut {NoStop}%
\bibitem [{\citenamefont {Iveland}\ \emph {et~al.}(2013)\citenamefont
  {Iveland}, \citenamefont {Martinelli}, \citenamefont {Peretti}, \citenamefont
  {Speck},\ and\ \citenamefont {Weisbuch}}]{Iveland2013}%
  \BibitemOpen
  \bibfield  {author} {\bibinfo {author} {\bibfnamefont {J.}~\bibnamefont
  {Iveland}}, \bibinfo {author} {\bibfnamefont {L.}~\bibnamefont {Martinelli}},
  \bibinfo {author} {\bibfnamefont {J.}~\bibnamefont {Peretti}}, \bibinfo
  {author} {\bibfnamefont {J.~S.}\ \bibnamefont {Speck}}, \ and\ \bibinfo
  {author} {\bibfnamefont {C.}~\bibnamefont {Weisbuch}},\ }\href {\doibase
  10.1103/physrevlett.110.177406} {\bibfield  {journal} {\bibinfo  {journal}
  {Physical Review Letters}\ }\textbf {\bibinfo {volume} {110}} (\bibinfo
  {year} {2013}),\ 10.1103/physrevlett.110.177406}\BibitemShut {NoStop}%
\bibitem [{\citenamefont {Schwartz}\ \emph {et~al.}(2012)\citenamefont
  {Schwartz}, \citenamefont {Naka}, \citenamefont {Kieseling},\ and\
  \citenamefont {Stolz}}]{Schwartz2012}%
  \BibitemOpen
  \bibfield  {author} {\bibinfo {author} {\bibfnamefont {R.}~\bibnamefont
  {Schwartz}}, \bibinfo {author} {\bibfnamefont {N.}~\bibnamefont {Naka}},
  \bibinfo {author} {\bibfnamefont {F.}~\bibnamefont {Kieseling}}, \ and\
  \bibinfo {author} {\bibfnamefont {H.}~\bibnamefont {Stolz}},\ }\href
  {\doibase 10.1088/1367-2630/14/2/023054} {\bibfield  {journal} {\bibinfo
  {journal} {New Journal of Physics}\ }\textbf {\bibinfo {volume} {14}},\
  \bibinfo {pages} {023054} (\bibinfo {year} {2012})}\BibitemShut {NoStop}%
\bibitem [{\citenamefont {Kumar}\ \emph {et~al.}(2014)\citenamefont {Kumar},
  \citenamefont {Cui}, \citenamefont {Ceballos}, \citenamefont {He},
  \citenamefont {Wang},\ and\ \citenamefont {Zhao}}]{Kumar2014}%
  \BibitemOpen
  \bibfield  {author} {\bibinfo {author} {\bibfnamefont {N.}~\bibnamefont
  {Kumar}}, \bibinfo {author} {\bibfnamefont {Q.}~\bibnamefont {Cui}}, \bibinfo
  {author} {\bibfnamefont {F.}~\bibnamefont {Ceballos}}, \bibinfo {author}
  {\bibfnamefont {D.}~\bibnamefont {He}}, \bibinfo {author} {\bibfnamefont
  {Y.}~\bibnamefont {Wang}}, \ and\ \bibinfo {author} {\bibfnamefont
  {H.}~\bibnamefont {Zhao}},\ }\href {\doibase 10.1103/physrevb.89.125427}
  {\bibfield  {journal} {\bibinfo  {journal} {Physical Review B}\ }\textbf
  {\bibinfo {volume} {89}} (\bibinfo {year} {2014}),\
  10.1103/physrevb.89.125427}\BibitemShut {NoStop}%
\bibitem [{\citenamefont {Yuan}\ and\ \citenamefont {Huang}(2015)}]{Yuan2015}%
  \BibitemOpen
  \bibfield  {author} {\bibinfo {author} {\bibfnamefont {L.}~\bibnamefont
  {Yuan}}\ and\ \bibinfo {author} {\bibfnamefont {L.}~\bibnamefont {Huang}},\
  }\href {\doibase 10.1039/c5nr00383k} {\bibfield  {journal} {\bibinfo
  {journal} {Nanoscale}\ }\textbf {\bibinfo {volume} {7}},\ \bibinfo {pages}
  {7402} (\bibinfo {year} {2015})}\BibitemShut {NoStop}%
\bibitem [{\citenamefont {Wang}\ \emph {et~al.}(2004)\citenamefont {Wang},
  \citenamefont {Dukovic}, \citenamefont {Knoesel}, \citenamefont {Brus},\ and\
  \citenamefont {Heinz}}]{Wang2004}%
  \BibitemOpen
  \bibfield  {author} {\bibinfo {author} {\bibfnamefont {F.}~\bibnamefont
  {Wang}}, \bibinfo {author} {\bibfnamefont {G.}~\bibnamefont {Dukovic}},
  \bibinfo {author} {\bibfnamefont {E.}~\bibnamefont {Knoesel}}, \bibinfo
  {author} {\bibfnamefont {L.~E.}\ \bibnamefont {Brus}}, \ and\ \bibinfo
  {author} {\bibfnamefont {T.~F.}\ \bibnamefont {Heinz}},\ }\href {\doibase
  10.1103/physrevb.70.241403} {\bibfield  {journal} {\bibinfo  {journal}
  {Physical Review B}\ }\textbf {\bibinfo {volume} {70}} (\bibinfo {year}
  {2004}),\ 10.1103/physrevb.70.241403}\BibitemShut {NoStop}%
\bibitem [{\citenamefont {Klimov}(2000)}]{Klimov2000}%
  \BibitemOpen
  \bibfield  {author} {\bibinfo {author} {\bibfnamefont {V.~I.}\ \bibnamefont
  {Klimov}},\ }\href {\doibase 10.1126/science.287.5455.1011} {\bibfield
  {journal} {\bibinfo  {journal} {Science}\ }\textbf {\bibinfo {volume}
  {287}},\ \bibinfo {pages} {1011} (\bibinfo {year} {2000})}\BibitemShut
  {NoStop}%
\bibitem [{\citenamefont {Taniguchi}\ and\ \citenamefont
  {Watanabe}(2007)}]{Taniguchi2007a}%
  \BibitemOpen
  \bibfield  {author} {\bibinfo {author} {\bibfnamefont {T.}~\bibnamefont
  {Taniguchi}}\ and\ \bibinfo {author} {\bibfnamefont {K.}~\bibnamefont
  {Watanabe}},\ }\href {\doibase 10.1016/j.jcrysgro.2006.12.061} {\bibfield
  {journal} {\bibinfo  {journal} {Journal of Crystal Growth}\ }\textbf
  {\bibinfo {volume} {303}},\ \bibinfo {pages} {525} (\bibinfo {year}
  {2007})}\BibitemShut {NoStop}%
\bibitem [{\citenamefont {Schu{\'{e}}}\ \emph {et~al.}(2016)\citenamefont
  {Schu{\'{e}}}, \citenamefont {Berini}, \citenamefont {Pla{\c{c}}ais},
  \citenamefont {Ducastelle}, \citenamefont {Barjon}, \citenamefont {Loiseau},\
  and\ \citenamefont {Betz}}]{Schue2016}%
  \BibitemOpen
  \bibfield  {author} {\bibinfo {author} {\bibfnamefont {L.}~\bibnamefont
  {Schu{\'{e}}}}, \bibinfo {author} {\bibfnamefont {B.}~\bibnamefont {Berini}},
  \bibinfo {author} {\bibfnamefont {B.}~\bibnamefont {Pla{\c{c}}ais}}, \bibinfo
  {author} {\bibfnamefont {F.}~\bibnamefont {Ducastelle}}, \bibinfo {author}
  {\bibfnamefont {J.}~\bibnamefont {Barjon}}, \bibinfo {author} {\bibfnamefont
  {A.}~\bibnamefont {Loiseau}}, \ and\ \bibinfo {author} {\bibfnamefont
  {A.}~\bibnamefont {Betz}},\ }\href {\doibase 10.1039/C6NR01253A} {\bibfield
  {journal} {\bibinfo  {journal} {Nanoscale}\ }\textbf {\bibinfo {volume}
  {8}},\ \bibinfo {pages} {6986} (\bibinfo {year} {2016})}\BibitemShut
  {NoStop}%
\bibitem [{\citenamefont {Casey}\ and\ \citenamefont
  {Jayson}(1971)}]{Casey1971}%
  \BibitemOpen
  \bibfield  {author} {\bibinfo {author} {\bibfnamefont {H.~C.}\ \bibnamefont
  {Casey}}\ and\ \bibinfo {author} {\bibfnamefont {J.~S.}\ \bibnamefont
  {Jayson}},\ }\href {\doibase 10.1063/1.1660623} {\bibfield  {journal}
  {\bibinfo  {journal} {Journal of Applied Physics}\ }\textbf {\bibinfo
  {volume} {42}},\ \bibinfo {pages} {2774} (\bibinfo {year}
  {1971})}\BibitemShut {NoStop}%
\bibitem [{\citenamefont {Cassabois}\ \emph {et~al.}(2016)\citenamefont
  {Cassabois}, \citenamefont {Valvin},\ and\ \citenamefont
  {Gil}}]{Cassabois2016}%
  \BibitemOpen
  \bibfield  {author} {\bibinfo {author} {\bibfnamefont {G.}~\bibnamefont
  {Cassabois}}, \bibinfo {author} {\bibfnamefont {P.}~\bibnamefont {Valvin}}, \
  and\ \bibinfo {author} {\bibfnamefont {B.}~\bibnamefont {Gil}},\ }\href
  {\doibase 10.1038/nphoton.2015.277} {\bibfield  {journal} {\bibinfo
  {journal} {Nat. Photonics}\ }\textbf {\bibinfo {volume} {10}},\ \bibinfo
  {pages} {262} (\bibinfo {year} {2016})}\BibitemShut {NoStop}%
\bibitem [{\citenamefont {Pierret}\ \emph {et~al.}(2014)\citenamefont
  {Pierret}, \citenamefont {Loayza}, \citenamefont {Berini}, \citenamefont
  {Betz}, \citenamefont {Pla\c{c}ais}, \citenamefont {Ducastelle},
  \citenamefont {Barjon},\ and\ \citenamefont {Loiseau}}]{Pierret2014}%
  \BibitemOpen
  \bibfield  {author} {\bibinfo {author} {\bibfnamefont {A.}~\bibnamefont
  {Pierret}}, \bibinfo {author} {\bibfnamefont {J.}~\bibnamefont {Loayza}},
  \bibinfo {author} {\bibfnamefont {B.}~\bibnamefont {Berini}}, \bibinfo
  {author} {\bibfnamefont {A.}~\bibnamefont {Betz}}, \bibinfo {author}
  {\bibfnamefont {B.}~\bibnamefont {Pla\c{c}ais}}, \bibinfo {author}
  {\bibfnamefont {F.}~\bibnamefont {Ducastelle}}, \bibinfo {author}
  {\bibfnamefont {J.}~\bibnamefont {Barjon}}, \ and\ \bibinfo {author}
  {\bibfnamefont {A.}~\bibnamefont {Loiseau}},\ }\href {\doibase
  10.1103/PhysRevB.89.035414} {\bibfield  {journal} {\bibinfo  {journal} {Phys.
  Rev. B}\ }\textbf {\bibinfo {volume} {89}},\ \bibinfo {pages} {035414}
  (\bibinfo {year} {2014})}\BibitemShut {NoStop}%
\bibitem [{\citenamefont {Watanabe}\ and\ \citenamefont
  {Taniguchi}(2011)}]{Watanabe2011}%
  \BibitemOpen
  \bibfield  {author} {\bibinfo {author} {\bibfnamefont {K.}~\bibnamefont
  {Watanabe}}\ and\ \bibinfo {author} {\bibfnamefont {T.}~\bibnamefont
  {Taniguchi}},\ }\href {\doibase 10.1111/j.1744-7402.2011.02626.x} {\bibfield
  {journal} {\bibinfo  {journal} {Int. J. Appl. Ceram. Technol.}\ }\textbf
  {\bibinfo {volume} {8}},\ \bibinfo {pages} {977} (\bibinfo {year}
  {2011})}\BibitemShut {NoStop}%
\bibitem [{\citenamefont {Cao}\ \emph {et~al.}(2013)\citenamefont {Cao},
  \citenamefont {Clubine}, \citenamefont {Edgar}, \citenamefont {Lin},\ and\
  \citenamefont {Jiang}}]{Cao2013}%
  \BibitemOpen
  \bibfield  {author} {\bibinfo {author} {\bibfnamefont {X.~K.}\ \bibnamefont
  {Cao}}, \bibinfo {author} {\bibfnamefont {B.}~\bibnamefont {Clubine}},
  \bibinfo {author} {\bibfnamefont {J.~H.}\ \bibnamefont {Edgar}}, \bibinfo
  {author} {\bibfnamefont {J.~Y.}\ \bibnamefont {Lin}}, \ and\ \bibinfo
  {author} {\bibfnamefont {H.~X.}\ \bibnamefont {Jiang}},\ }\href {\doibase
  10.1063/1.4829026} {\bibfield  {journal} {\bibinfo  {journal} {Appl. Phys.
  Lett.}\ }\textbf {\bibinfo {volume} {103}},\ \bibinfo {pages} {191106}
  (\bibinfo {year} {2013})}\BibitemShut {NoStop}%
\bibitem [{\citenamefont {{Paleari}}\ \emph {et~al.}(2018)\citenamefont
  {{Paleari}}, \citenamefont {{Galvani}}, \citenamefont {{Amara}},
  \citenamefont {{Ducastelle}}, \citenamefont {{Molina-S{\'a}nchez}},\ and\
  \citenamefont {{Wirtz}}}]{Paleari2018}%
  \BibitemOpen
  \bibfield  {author} {\bibinfo {author} {\bibfnamefont {F.}~\bibnamefont
  {{Paleari}}}, \bibinfo {author} {\bibfnamefont {T.}~\bibnamefont
  {{Galvani}}}, \bibinfo {author} {\bibfnamefont {H.}~\bibnamefont {{Amara}}},
  \bibinfo {author} {\bibfnamefont {F.}~\bibnamefont {{Ducastelle}}}, \bibinfo
  {author} {\bibfnamefont {A.}~\bibnamefont {{Molina-S{\'a}nchez}}}, \ and\
  \bibinfo {author} {\bibfnamefont {L.}~\bibnamefont {{Wirtz}}},\ }\href@noop
  {} {\bibfield  {journal} {\bibinfo  {journal} {ArXiv e-prints}\ } (\bibinfo
  {year} {2018})},\ \Eprint {http://arxiv.org/abs/1803.00982} {arXiv:1803.00982
  [cond-mat.mtrl-sci]} \BibitemShut {NoStop}%
\bibitem [{\citenamefont {Mouri}\ \emph {et~al.}(2014)\citenamefont {Mouri},
  \citenamefont {Miyauchi}, \citenamefont {Toh}, \citenamefont {Zhao},
  \citenamefont {Eda},\ and\ \citenamefont {Matsuda}}]{Mouri2014a}%
  \BibitemOpen
  \bibfield  {author} {\bibinfo {author} {\bibfnamefont {S.}~\bibnamefont
  {Mouri}}, \bibinfo {author} {\bibfnamefont {Y.}~\bibnamefont {Miyauchi}},
  \bibinfo {author} {\bibfnamefont {M.}~\bibnamefont {Toh}}, \bibinfo {author}
  {\bibfnamefont {W.}~\bibnamefont {Zhao}}, \bibinfo {author} {\bibfnamefont
  {G.}~\bibnamefont {Eda}}, \ and\ \bibinfo {author} {\bibfnamefont
  {K.}~\bibnamefont {Matsuda}},\ }\href {\doibase 10.1103/physrevb.90.155449}
  {\bibfield  {journal} {\bibinfo  {journal} {Physical Review B}\ }\textbf
  {\bibinfo {volume} {90}} (\bibinfo {year} {2014}),\
  10.1103/physrevb.90.155449}\BibitemShut {NoStop}%
\bibitem [{\citenamefont {Surrente}\ \emph {et~al.}(2016)\citenamefont
  {Surrente}, \citenamefont {Mitioglu}, \citenamefont {Galkowski},
  \citenamefont {Klopotowski}, \citenamefont {Tabis}, \citenamefont {Vignolle},
  \citenamefont {Maude},\ and\ \citenamefont {Plochocka}}]{Surrente2016}%
  \BibitemOpen
  \bibfield  {author} {\bibinfo {author} {\bibfnamefont {A.}~\bibnamefont
  {Surrente}}, \bibinfo {author} {\bibfnamefont {A.~A.}\ \bibnamefont
  {Mitioglu}}, \bibinfo {author} {\bibfnamefont {K.}~\bibnamefont {Galkowski}},
  \bibinfo {author} {\bibfnamefont {L.}~\bibnamefont {Klopotowski}}, \bibinfo
  {author} {\bibfnamefont {W.}~\bibnamefont {Tabis}}, \bibinfo {author}
  {\bibfnamefont {B.}~\bibnamefont {Vignolle}}, \bibinfo {author}
  {\bibfnamefont {D.~K.}\ \bibnamefont {Maude}}, \ and\ \bibinfo {author}
  {\bibfnamefont {P.}~\bibnamefont {Plochocka}},\ }\href {\doibase
  10.1103/physrevb.94.075425} {\bibfield  {journal} {\bibinfo  {journal}
  {Physical Review B}\ }\textbf {\bibinfo {volume} {94}} (\bibinfo {year}
  {2016}),\ 10.1103/physrevb.94.075425}\BibitemShut {NoStop}%
\bibitem [{\citenamefont {Engel}\ \emph {et~al.}(2006)\citenamefont {Engel},
  \citenamefont {Leo},\ and\ \citenamefont {Hoffmann}}]{Engel2006}%
  \BibitemOpen
  \bibfield  {author} {\bibinfo {author} {\bibfnamefont {E.}~\bibnamefont
  {Engel}}, \bibinfo {author} {\bibfnamefont {K.}~\bibnamefont {Leo}}, \ and\
  \bibinfo {author} {\bibfnamefont {M.}~\bibnamefont {Hoffmann}},\ }\href
  {\doibase 10.1016/j.chemphys.2005.09.004} {\bibfield  {journal} {\bibinfo
  {journal} {Chemical Physics}\ }\textbf {\bibinfo {volume} {325}},\ \bibinfo
  {pages} {170} (\bibinfo {year} {2006})}\BibitemShut {NoStop}%
\bibitem [{\citenamefont {Ma}\ \emph {et~al.}(2012)\citenamefont {Ma},
  \citenamefont {Xiao},\ and\ \citenamefont {Shaw}}]{Ma2012}%
  \BibitemOpen
  \bibfield  {author} {\bibinfo {author} {\bibfnamefont {Y.-Z.}\ \bibnamefont
  {Ma}}, \bibinfo {author} {\bibfnamefont {K.}~\bibnamefont {Xiao}}, \ and\
  \bibinfo {author} {\bibfnamefont {R.~W.}\ \bibnamefont {Shaw}},\ }\href
  {\doibase 10.1021/jp3057543} {\bibfield  {journal} {\bibinfo  {journal} {The
  Journal of Physical Chemistry C}\ }\textbf {\bibinfo {volume} {116}},\
  \bibinfo {pages} {21588} (\bibinfo {year} {2012})}\BibitemShut {NoStop}%
\bibitem [{\citenamefont {Shaw}\ \emph {et~al.}(2008)\citenamefont {Shaw},
  \citenamefont {Ruseckas},\ and\ \citenamefont {Samuel}}]{Shaw2008}%
  \BibitemOpen
  \bibfield  {author} {\bibinfo {author} {\bibfnamefont {P.~E.}\ \bibnamefont
  {Shaw}}, \bibinfo {author} {\bibfnamefont {A.}~\bibnamefont {Ruseckas}}, \
  and\ \bibinfo {author} {\bibfnamefont {I.~D.~W.}\ \bibnamefont {Samuel}},\
  }\href {\doibase 10.1002/adma.200800982} {\bibfield  {journal} {\bibinfo
  {journal} {Advanced Materials}\ }\textbf {\bibinfo {volume} {20}},\ \bibinfo
  {pages} {3516} (\bibinfo {year} {2008})}\BibitemShut {NoStop}%
\bibitem [{\citenamefont {Yoshioka}\ \emph {et~al.}(2010)\citenamefont
  {Yoshioka}, \citenamefont {Ideguchi}, \citenamefont {Mysyrowicz},\ and\
  \citenamefont {Kuwata-Gonokami}}]{Yoshioka2010}%
  \BibitemOpen
  \bibfield  {author} {\bibinfo {author} {\bibfnamefont {K.}~\bibnamefont
  {Yoshioka}}, \bibinfo {author} {\bibfnamefont {T.}~\bibnamefont {Ideguchi}},
  \bibinfo {author} {\bibfnamefont {A.}~\bibnamefont {Mysyrowicz}}, \ and\
  \bibinfo {author} {\bibfnamefont {M.}~\bibnamefont {Kuwata-Gonokami}},\
  }\href {\doibase 10.1103/physrevb.82.041201} {\bibfield  {journal} {\bibinfo
  {journal} {Physical Review B}\ }\textbf {\bibinfo {volume} {82}} (\bibinfo
  {year} {2010}),\ 10.1103/physrevb.82.041201}\BibitemShut {NoStop}%
\bibitem [{\citenamefont {Snoke}\ and\ \citenamefont
  {Kavoulakis}(2014)}]{Snoke2014}%
  \BibitemOpen
  \bibfield  {author} {\bibinfo {author} {\bibfnamefont {D.}~\bibnamefont
  {Snoke}}\ and\ \bibinfo {author} {\bibfnamefont {G.~M.}\ \bibnamefont
  {Kavoulakis}},\ }\href {\doibase 10.1088/0034-4885/77/11/116501} {\bibfield
  {journal} {\bibinfo  {journal} {Reports on Progress in Physics}\ }\textbf
  {\bibinfo {volume} {77}},\ \bibinfo {pages} {116501} (\bibinfo {year}
  {2014})}\BibitemShut {NoStop}%
\bibitem [{\citenamefont {Watanabe}\ \emph {et~al.}(2009)\citenamefont
  {Watanabe}, \citenamefont {Taniguchi}, \citenamefont {Niiyama}, \citenamefont
  {Miya},\ and\ \citenamefont {Taniguchi}}]{Watanabe2009a}%
  \BibitemOpen
  \bibfield  {author} {\bibinfo {author} {\bibfnamefont {K.}~\bibnamefont
  {Watanabe}}, \bibinfo {author} {\bibfnamefont {T.}~\bibnamefont {Taniguchi}},
  \bibinfo {author} {\bibfnamefont {T.}~\bibnamefont {Niiyama}}, \bibinfo
  {author} {\bibfnamefont {K.}~\bibnamefont {Miya}}, \ and\ \bibinfo {author}
  {\bibfnamefont {M.}~\bibnamefont {Taniguchi}},\ }\href {\doibase
  10.1038/NPHOTON.2009.167} {\bibfield  {journal} {\bibinfo  {journal} {Nat.
  Photonics}\ }\textbf {\bibinfo {volume} {3}},\ \bibinfo {pages} {591}
  (\bibinfo {year} {2009})}\BibitemShut {NoStop}%
\bibitem [{\citenamefont {Wang}\ \emph {et~al.}(2006)\citenamefont {Wang},
  \citenamefont {Wu}, \citenamefont {Hybertsen},\ and\ \citenamefont
  {Heinz}}]{Wang2006}%
  \BibitemOpen
  \bibfield  {author} {\bibinfo {author} {\bibfnamefont {F.}~\bibnamefont
  {Wang}}, \bibinfo {author} {\bibfnamefont {Y.}~\bibnamefont {Wu}}, \bibinfo
  {author} {\bibfnamefont {M.~S.}\ \bibnamefont {Hybertsen}}, \ and\ \bibinfo
  {author} {\bibfnamefont {T.~F.}\ \bibnamefont {Heinz}},\ }\href {\doibase
  10.1103/physrevb.73.245424} {\bibfield  {journal} {\bibinfo  {journal}
  {Physical Review B}\ }\textbf {\bibinfo {volume} {73}} (\bibinfo {year}
  {2006}),\ 10.1103/physrevb.73.245424}\BibitemShut {NoStop}%
\bibitem [{\citenamefont {Wirtz}\ \emph {et~al.}(2006)\citenamefont {Wirtz},
  \citenamefont {Marini},\ and\ \citenamefont {Rubio}}]{Wirtz2006}%
  \BibitemOpen
  \bibfield  {author} {\bibinfo {author} {\bibfnamefont {L.}~\bibnamefont
  {Wirtz}}, \bibinfo {author} {\bibfnamefont {A.}~\bibnamefont {Marini}}, \
  and\ \bibinfo {author} {\bibfnamefont {A.}~\bibnamefont {Rubio}},\ }\href
  {\doibase 10.1103/PhysRevLett.96.126104} {\bibfield  {journal} {\bibinfo
  {journal} {Phys. Rev. Lett.}\ }\textbf {\bibinfo {volume} {96}},\ \bibinfo
  {pages} {126104} (\bibinfo {year} {2006})}\BibitemShut {NoStop}%
\end{thebibliography}%

\end{document}